\documentstyle[]{article}

\def\ltsima{$\; \buildrel < \over \sim \;$}
\def\lsim{\lower.5ex\hbox{\ltsima}}
\def\gtsima{$\; \buildrel > \over \sim \;$}
\def\gsim{\lower.5ex\hbox{\gtsima}}

\def\aa #1 #2 {A\&A, {#1}, #2}
\def\aas #1 #2 {A\&AS, {#1}, #2}
\def\araa #1 #2 {ARA\&A, {#1}, #2}
\def\mon #1 #2 {MNRAS, {#1}, #2}
\def\apj #1 #2 {ApJ, {#1}, #2}
\def\apjs #1 #2 {ApJS, {#1}, #2}
\def\apjl #1 #2 {ApJ Lett, {#1}, #2}
\def\aj #1 #2 {AJ, {#1}, #2}
\def\nat #1 #2 {Nature, {#1}, #2}
\def\pasj #1 #2 {PASJ, {#1}, #2}
\def\pasp #1 #2 {PASP, {#1}, #2}
\newcommand{\gras}[1]{\mbox{\boldmath $#1$}}

\begin{document}

\begin{center}

\begin{huge}
{\bf The pulsed soft X--ray emission from PSR~0656+14}
\end{huge}

\vskip 1.0cm

{\bf A.~Possenti}
  
{\small {\it Dipartimento di Fisica, Universit\`a di Milano, 
via Celoria 16, I-20133 Milano, Italy}}

{\bf S.~Mereghetti}

{\small {\it Istituto di Fisica Cosmica del C.N.R., 
Via Bassini 15, I-20133 Milano, 
Italy}}

{\bf M.~Colpi}

{\small {\it Dipartimento di Fisica, Universit\`a di Milano, 
via Celoria 16, I-20133 Milano, Italy}}

\end{center}

\vskip 0.5cm

\begin{small}
Send offprint requests to: {\it S.~Mereghetti;~e-mail:~sandro@ifctr.mi.cnr.it}

Accepted by Astronomy \& Astrophysics on 29 February 1996
\end{small}

\vskip 0.5cm

\begin{abstract}

We present the results of a spectral and timing analysis of
PSR~0656+14 based on the complete set of ROSAT observations carried
out with the PSPC instrument in 1991 and 1992.  The present  analysis
confirms the thermal origin of the bulk of the  emission in the soft
X-ray band (Finley et al. 1992).  In addition, we find  strong
evidence of a harder component, described equally well with a
blackbody  at $T \simeq 2\times10^6$ K, or with a steep  power law
with photon index $\Gamma \simeq 4.5.$ This bimodal emission is also
supported by an analysis of the light curve shape as a function of the
energy. 

The  0.1--2.4 keV light curve of PSR~0656+14, with a pulsed fraction
of about 9\%, is interpreted with a simple model for the temperature
di\-stri\-bution on the neutron star surface, coupled with the geometrical
information derived from radio data. In this model, which includes the
effects of relativistic light bending and gravitational redshift, the
X--rays ori\-gi\-na\-te from two thermal components resulting from neutron
star cooling and blackbody emission released in the hotter polar  cap
regions. 
 
The observed modulation can be reproduced only if PSR~0656+14 has a
relatively high dipole inclination ($\sim30^o$) and $(1+z)~\lsim$1.15.
The overall pulsed fraction cannot be significantly increased by
including the polar cap contribution, if its temperature and intensity
are to be consistent with the observed spectra. 
    
\end{abstract}
   
\section{Introduction}

ROSAT ob\-ser\-va\-tions of isolated neutron stars with characteristic
ages of  a few $10^{5}$ years (PSR~0656+14, PSR~1055-52, Geminga) have
shown that the bulk of their X--ray e\-mis\-sion is thermal in origin
and may result from neutron star cooling (see \"Ogelman 1995 for a
review). The observed periodic modulation in the X--ray flux indicates
the presence of large thermal gradients on the surface of these
magnetized neutron stars. For a given surface temperature
distribution, the degree of modulation depends on the intrinsic
geometry of the source and on its orientation relative to the
observer, which in the case of radio pulsars can be inferred from the
shape, polarization and spectrum of the radio pulses. 

PSR~0656+14, first studied in the  soft X--ray range with the 
Einstein Observatory (Cordova et al. 1989), is the brightest 
''middle-aged''   neutron star      in this energy band. Its  
association with a  possible X--ray supernova remnant, the
Gemini--Monoceros ring (Nousek et al. 1981; Thompson et al. 1991) has
recently been criticized, on the basis of new results on the proper
motion of the pulsar (Thompson \& Cordova 1994). A candidate optical
counterpart with V$\sim25$ has been discovered by Caraveo, Bignami \&
Mereghetti (1994) within 1'' of the radio position of PSR~0656+14. 

Three   pointings of PSR~0656+14 were carried out with the ROSAT PSPC
detector (1991 March 26, 1992 March 24 and April 15). While the first
observation ($\sim$3200 s) has been presented in detail by Finley,
\"Ogelman \& Kizilo\u{g}lu (1992), only the preliminary results of the
two 1992 pointings have been reported (\"Ogelman 1995).

In this paper we analyze all the PSPC data of PSR~0656+14 and
interpret them in the context of a simple model for its temperature
distribution  coupled with the geometrical configuration derived from
the radio data. On the basis of the new observational results, the
paper carries out an analysis similar to the one presented by Page
(1995a) on PSR~0656+14 and explores  the effect of polar cap emission
on the light curves and spectra. 

\section{ROSAT observations: Data analysis and results}

Our spectral analysis is based on the sum of the three observations,
for a total exposure time of about 17000 s. The source counts were
extracted from a circular region (3' radius) centered at the position
of PSR~0656+14. The background was estimated from a concentric annular
region with radii 4' and 10' (due to the high source count rate of
$\sim 1.9$ ct s$^{-1}$, the results are not significantly dependent on
the particular region used for the background determination). The
selected counts ($\sim 32,000$, in channels  11--240) were rebinned in
order to achieve a minimum signal to noise ratio of 7 in each  energy
channel and then fitted to different model spectra. 

Single component models (power law, blackbody, thermal bremsstrahlung)
gave unacceptable results. Figure 1 shows the fit residuals obtained
in the case of a blackbody (the model giving the lowest $\chi^2$).
While such a model was still compatible with the data of the first
observation alone (Finley et al. 1992), the better statistics of the
complete data set clearly require  the presence of a second, harder
component dominating the X--ray emission above $\sim$ 1~keV. We
therefore considered composite models, consisting of a ``soft''
blackbody and a ``hard tail'' in the form of either a second blackbody
or a power law. 

For the former case (see Figure~2a) a good fit  ($\chi^2$=1.07 for
74~dof) was obtained with blackbody temperatures $T_{1}^{\infty}=$9.1
$\times10^{5}~{}^oK$ and $T_{2}^{\infty}=$1.9$\times10^{6}~{}^oK$ for
the soft and hard components, respectively, and an interstellar
absorption $N_{H}=$6.9$\times 10^{19}$ cm$^{-2}.$ Figure~3 (a,b) gives
the confidence contours for these quantities, while Table~1 summarizes
the best fit values of all the parameters. The flux in the soft
component implies an emitting region with radius $R^{\infty}\sim$14 km
(for a distance of 760~pc; Taylor et al. 1995). For the same distance,
the total luminosity is about $10^{33}~$erg s$^{-1}.$ Figure~2b shows
the best fit with a blackbody plus power law, corresponding to a
reduced $\chi^2$=1.05. The best fit photon index is $\Gamma=4.5.$
Contrary to the previous case, in this spectral decomposition both
components have a similar flux. In the fit of the hard tail with a
power law, one obtains values for $N_H$ and for the soft balckbody
temperature significantly higher than those of the two blackbody fit
(see Figures~3c and 3d).

%%%%%%%%%%%%%%% Table 1 %%%%%%%%%%%%%%%%%%%%%%%%%%%%%%%%%%%%%%%%%%%%%%%
\begin{table}[tbh]
\begin{center}
\caption{Results of the spectral   fit of PSR~0656+14}
\label{tab:fit}
\vspace{0.0cm}
\begin{tabular}{c||c|ccc|c}
\hline
\hline
     Model       &        $N_{H}$       &      T$^{\infty}(10^6~{}^oK)$
           &       Flux                & $R^{\infty}$ &$\chi^2/dof$
\\
                 & (10$^{20}$cm$^{-2}$) &      or photon index $\Gamma$
           & (erg cm$^{-2}$sec$^{-1})$ &      (km)    &
\\
\hline
\hline
Blackbody (soft) &                      &              0.91$~\pm~$0.05
            &   1.3$~\times 10^{-11}$   &      14.0    &
\\
       +         &    0.7$~\pm~$0.1     &                             
            &                           &              &    1.07
\\
Blackbody (hard) &                      &              1.9$~\pm~$0.4   
            &   9.4$~\times 10^{-13}$   &      0.8     &
\\
\hline
Blackbody (soft) &                      &              1.01$~\pm~$0.06
            &   9.1$~\times 10^{-12}$   &      9.1     &
\\
       +         &    1.9$~\pm~$0.4     &                             
            &                           &              &    1.05
\\
Power-law        &                      &              4.5$~\pm~$0.4  
            &   4.1$~\times 10^{-11}$   &       ---    &
\\
\hline
\hline
\end{tabular}
\end{center}
Errors are at 90\% confidence level for a single interesting
parameter.  $R^{\infty}$ is for a distance d=760 pc. The fluxes are 
unabsorbed in the 0.1--2.4 keV range.
\end{table}
%%%%%%%%%%%% End Table 1 %%%%%%%%%%%%%%%%%%%%%%%%%%%%%%%%%%%%%%%%%%%%%%%

For the timing analysis, we only considered the two 1992 observations,
since the relatively short time gap between them allows an accurate
relative measure of the photon arrival times. In order to maximize the
signal to noise ratio, the source counts were extracted from circles
with different radii (160" --- 50"), matched to the PSPC point
response function at the different energies. After the correction of
the arrival times to the Solar System Barycenter, we folded the data
using the $P$ and $\dot{P}$ values expected from the radio
observations (Taylor et al. 1995), to obtain the 0.1 -- 2.4 keV light
curve   shown in Figure~4 (we also verified that a standard  period
search  resulted in a best value for $P$ compatible with that expected
{}from the radio data). The light curve shows a single broad peak,
similar to that of the 1991 data (Finley et al. 1992), but with a
slightly smaller  pulsed fraction of  $P_{fr}=9\%$ (
$P_{fr}$=(CR$_{ave}$-CR$_{min}$)/CR$_{ave}$, where CR$_{ave}$ and
CR$_{min}$ are the mean and minimum count rates). As shown in
Figure~5, the degree of modulation and the phase of the peak depend on
the photon energy: above $\sim~$0.5 keV the peak is narrower and
shifted of about 70$^o$. Contrary to the case of Geminga (Halpern \&
Ruderman 1993), there is no evidence of an anticorrelation between the
pulsed fraction and the photon energy in the soft ($\lsim~$0.5 keV)
component (see also Page, Shibanov \& Zavlin 1995). The different
shape and higher pulsed fraction above 0.5 keV in the light curves of
PSR~0656+14, support the presence of two different components
suggested by the spectral analysis.

\section{Simulation of pulsed light curves}

\subsection{The model}

Our model consists of the superposition of two contributions: a
thermal component with nonuniform temperature distribution which
arises from the whole surface, and an additional emission from hotter
region(s) near the magnetic poles. The first component is modelled
following the work by Greenstein \& Hartke (1983), who derived the 
surface temperature gradient resulting from anisotropic heat transport
{}from the hotter isothermal interior to the cooler surface. The degree
of anisotropy of the heat flow, induced by a crustal magnetic field,
is parameterized by the ratio of the coefficients of thermal
conductivity orthogonal and parallel to the field:
$K=k_{\perp}/k_{\parallel}$. In the hypothesis of a dipolar field, the
surface temperature reaches a maximum at the magnetic poles (where the
field is almost orthogonal to the surface) and a minimum along the
magnetic equator. For middle aged neutron stars, such as PSR~0656+14,
$K$ is expected to lie  in the range 0.0001 --- 0.01 (Schaaf 1990;
Hernquist  1985). Since the temperature distribution depends only
weakly on $K$, the  light curve shape  is only slightly influenced by
this parameter. We have verified that our results are independent of
the   exact value of $K$ within this range and then    adopted
$K$=0.001. 

It is possible to derive a relation between the surface temperature
$T_{surf}$ and the magnetic colatitude $\chi$ ($\chi=0$ at the
magnetic poles): 

$$T_{surf}(\chi)=T_{eff}  \Bigl [ 1 - 0.47(1-K) \Bigr ] ^{-1/4} 
\Bigl [ \frac{K+(4-K) \cos^2 \chi}{1+3\cos^2 \chi} \Bigr ] 
^{1/4}\eqno{(1)}$$

\noindent
where $T_{eff}$ is the effective temperature of the star, i.e. the
temperature of the blackbody with the same bolometric luminosity  of a
source with a temperature distribution given by (1). In the following,
we have estimated $T_{eff}$ on the basis of the results of blackbody
spectral fits. In fact, the spectra produced by (1) cannot be
distinguished from that of a blackbody at $T_{eff}$ within the limited
energy resolution and statistics of the present observations. Due to
the effects of the radiative transport in the neutron star atmosphere
on the emerging spectra, the surface temperature can differ from that
derived with pure blackbody fits (Romani 1987, Shibanov et al. 1992).
However, the exact value of $T_{eff}$ adopted in the computations has
little effect on the degree of modulation of the resulting light
curves, which mainly depend on orientation and gravitational effects.
Since our results would remain practically unchanged for any other
locally isotropic emission spectrum, we have adopted for simplicity
the blackbody model.

The magnetosphere of a neutron star is characterized by strong
magnetic fields, high-energy photons and charged particles of opposite
polarities. It is not surprising that, under favourable conditions, a
fraction among the charged particles is accelerated towards the star
(see, e.g. Arons 1981; Cheng, Ho \& Ruderman 1986a, 1986b), colliding
with it in the polar-cap regions. The kinetic energy of the impinging
particles produces a reheating of the caps, which become hotter than
the contiguous regions. To account for this phenomenon, we introduce
two hot areas near the  magnetic poles: their temperature,
$T_{cap}^{\infty}$, and   angular size, $\theta_{cap}$, are determined
{}from the results relative to the hard tail. 

This relatively simple model applies if the surface magnetic field is
the primary factor in determining the temperature distribution: it
becomes invalid if the thermal structure at the neutron star surface
is instead controlled by irradiation of hard X-rays backscattered in
the magnetosphere (Halpern \& Ruderman 1993). Moreover, absorption and
scattering of radiation in the magnetospheric plasma can alter the
pulse profiles and weaken the effect of an anisotropic surface
temperature distribution (Zavlin, Shibanov \& Pavlov 1995). These
effects are not included.

Since the soft X--ray emission originates at the neutron star's
surface, where the gravitational field is very strong, the propagation
of this radiation must be examined in the framework of general
relativity. The two outstanding effects (Page 1995a) are the change of
the energy of the photons along their path (gra\-vi\-ta\-tional red-shift)
and the bending of their trajectories (gravitational bending). We
include both effects in the calculation of the light curves and
spectra, using the method outlined by Pechenick et al. (1983). The
physical parameter describing the extent of the relativistic effects
is $~(~1~+~z~)~=~(~1~-~2GM/Rc^2~)^{-1/2}$. 

After applying these corrections, the   light curves are derived by
integrating the contributions of  all the surface elements visible at
a given phase. The resulting emission is then  folded through the 
interstellar absorption and the instrumental response. 
 
The modulation of the  light curves depends on the angles between the 
spin and magnetic axes ($\alpha$) and between the latter and the line
of sight ($\beta$). We have considered the values derived for
PSR~0656+14 by radio observations. Two different approaches (Lyne \&
Manchester 1988; Rankin 1993) have been used to interpret the variety
of pulse shapes, po\-la\-ri\-za\-tion patterns and spectra of radio
pulsars. For several objects these models lead to significantly
different results  (see Miller \& Hamilton 1993, for a  comparison of
the two methods). For PSR~0656+14, Lyne \& Manchester derived
$\alpha=\beta=~8.2^o$, while  Rankin only gave an estimate for
$\alpha~=~30^o$, similar to the value ($\alpha~=~35^o$) independently
obtained by Malov (1990). 

%%%%%%%%%%%%%%% Table 2 %%%%%%%%%%%%%%%%%%%%%%%%%%%%%%%%%%%%%%%%%%%%%%%%
\begin{table}[tbh]
\begin{center}
\caption{Pulsed fractions of the simulated light curves (0.1--2.4 keV)
as a function of the adopted pa\-ra\-me\-ters. 
\label{tab:integ_par}}
\vspace{0.2cm}
\begin{tabular}{c||c|c|c}
\hline
parameters  &  & reference values &  
\\ 
\hline
\hline
$N_H$ (cm$^{-2}$)            & $~~~~1.7\cdot 10^{19}~~~~$ & 
$\gras{6.86\cdot 10^{19}}$ & $~~~~23\cdot 10^{19}~~~~$  
\\ 
                             &          $7.06\%$           &
          $7.46\%$          &          $8.35\%$      
\\
\hline
$T_{eff}^{\infty}~(^oK)$     &     $0.50\cdot 10^{6}$      &
$\gras{0.91\cdot 10^{6}}$  &   $1.30\cdot 10^{6}$     
\\
                             &          $8.81\%$           &
          $7.46\%$          &          $7.10\%$       
\\
\hline
$K$                          &          $0.0001$           &
       $\gras{0.0012}$      &           $0.01$         
\\ 
                             &          $7.46\%$           &
          $7.46\%$          &          $7.36\%$      
\\
\hline
$\theta_{cap}$               &           $0^o$             &
        $\gras{3.4^o}$      &             $$         
\\ 
                             &          $7.00\%$           &
          $7.46\%$          &             $$       
\\
\hline
$T_{cap}^{\infty}~(^oK)$     &    $1.50\cdot 10^{6}$       &
$\gras{1.90\cdot 10^{6}}$  &   $2.30\cdot 10^{6}$    
\\ 
                             &          $7.30\%$           &
          $7.46\%$          &          $7.76\%$        
\\
\hline
$\beta$                      &          $-10^o$            &
       $\gras{0^o}$         &          $+10^o$
\\ 
                             &          $5.56\%$           &
          $7.46\%$          &          $7.65\%$    
\\
\hline
\hline
$1+z$                        &     $1.10$                  &
       $\gras{1.15}$        &          $1.20$     
\\ 
                             &     $9.87\%$                &
          $7.46\%$          &          $5.15\%$      
\\
\hline
$\alpha$                     &           $8.2^o$           &
       $\gras{30.0^o}$      &          $35.0^o$        
\\ 
                             &          $L.\&M.$           &
          $Rankin$          &          $Malov$          
\\
                             &          $1.46\%$           &
          $7.46\%$          &          $8.55\%$         
\\
\hline
\hline
\end{tabular}
\end{center}
The adopted parameters of reference are shown in the central column.
Next to them, we report a set of values spanning the reliable range of
variability for each quantity. The pulsed fractions indicated below
each number are those obtained by setting all the other parameters at
the reference values. The observed pulsed fraction is in the range
[8--10\%]. 
\end{table}
%%%%%%%%%%%% End Table 2 %%%%%%%%%%%%%%%%%%%%%%%%%%%%%%%%%%%%%%%%%%%%%%

\subsection{Results}

Based on the above model and using the values of
$N_{H},~T_{eff}^{\infty},~T_{cap}^{\infty},~\theta_{cap}$ obtained
{}from the spectral fits of $\S 2$, we have simulated various light
curves. $T_{eff}$ and $~T_{cap}$ are fixed by choosing different
values of $(1+z)$ and the corresponding light curves are compared to
the observations. 
 
If PSR 0656+14 is in the geometrical configuration proposed by Rankin
($\alpha~=~30^o$), the calculated light curves can satisfactorily
reproduce the observed degree of modulation only if 1.10$~\le (1+z)
\le~$1.15. Figure~4 compares the observed 0.1--2.4  keV  light curve
to that computed with the values of reference shown in the central
column of Table~\ref{tab:integ_par}, giving a pulsed fraction
$P_{fr}=7.5\%.$ If $(1+z)$ is increased, $P_{fr}$ reduces
significantly and the resulting light curves are inconsistent with the
observation. As shown in Table~\ref{tab:integ_par}, $P_{fr}$ is little
affected by the values of the remaining parameters, except for $\beta$
(which cannot be much larger than 10$^o$ otherwise the pulsar would
not  be visible in the radio band). The emission from the polar caps
gives only a modest increase to the pulsed fraction in the 0.1 -- 2.4
keV range, dominated by the contribution of the softer blackbody. 

Following the same procedure, we have computed the light curves for
the angles suggested by Lyne \& Manchester
($\alpha=8.2^o,~\beta=8.2^o$). Our model cannot explain the relatively
high observed modulation, even for the unacceptable value of
$(1+z)=$1.00$~$. Using the combination of parameters which maximizes
the modulation, the predicted $P_{fr}$ would reach  at most 4$\%$,
much less than the minimum value compatible with the observations.
Thus, as also indicated by Page (1995a),   the small dipole
inclination angle proposed by Lyne \& Man\-che\-ster (1988) cannot account
for the observed pulsed fraction (see also Possenti, Mereghetti \&
Colpi 1996). 

\section{Conclusions}  

The ROSAT spectral and timing analysis of PSR~0656+14 indicate that
its soft X--ray emission  is characterized by two separate components.
The blackbody  temperature of the  softer component
($9\times10^5~^o$K), independent of the model used for the hard tail, 
is similar to that derived by Finley et al. (1992). 

The hard tail can be described equally well by a blackbody or by a
steep power law. In the blackbody case, it is natural to attribute the
hard component to localized emission from hotter regions of the star
surface, e.g. the polar caps, as discussed above. In the case of a
power law, the tail can be interpreted as a non-thermal emission of
magnetospheric origin. The best fit power law index ($\Gamma =4.5$) is
greater than that observed in  PSR~1055--52 ($\Gamma =1.5$ , \"Ogelman
\& Finley 1993) and Geminga ($\Gamma =2.5$, Halpern \& Ruderman 1993),
the other   isolated neutron stars of comparable age and  with a
similar two component soft X--ray emission. These two objects have
also been clearly observed at $\gamma$-ray energies (E$\gsim$100 MeV),
contrary to PSR~0656+14 for which only a marginal detection, at a much
lower flux, has been claimed (Ramanamurthy et al. 1996). This fact 
might indicate a possible connection between the level of 
$\gamma$--ray emission and the slope of the hard X--ray tail. 

With a model for the anisotropic thermal cooling induced by the
crustal, dipolar magnetic field (similar to our first component), Page
(1995a)  derived for PSR~0656+14 pulsed fractions between 8 and 10\%,
smaller than that ($\simeq$14\%) reported by Finley et al. (1992).
This result is confirmed by our analysis of the complete PSPC data 
set. We show that, even in the most favourable geo\-me\-tric configuration
($\alpha \sim 30^o$) among those compatible with the radio data, the
observed modulation can be reproduced only for $(1+z)\le$1.15
(corresponding to $R/R_{S}~\gsim$4, where $R_{S}$ is the Schwarzschild
radius). These low values of $z$ are consistent with a stellar radius
between $12$ and $17$ km and a neutron star mass $M\sim 1 M_{\odot}$.
For M = 1.4 $M_{\odot}$, our limit on $z$ requires a radius  $\sim 17$
km, a value     inconsistent with the most recent  neutron star models
(Glendenning 1985; Wiringa et al. 1988). Masses below 1.4 $M_{\odot}$
have radii sufficiently large to account for the observed modulation
in the X-ray signal, only  if the equation of state is very stiff
(Page 1995a). 

The presence of higher order moments in the magnetic field, or the
effects of a magnetized atmosphere causing anisotropic photon
transport have been proposed as solutions for increasing the light
curve modulation (Page 1995a, 1995b). Our analysis shows that the
alternative possibility of obtaining a higher modulation by including
the additional contribution of hot   polar caps does not solve the
problem. The polar cap emission has little effect on the overall
mo\-du\-la\-tion, if its  temperature and intensity are to be consistent
with the observed spectra.

\newpage

\begin{Large}

\noindent
{\bf Captions}

\end{Large}

\vskip 1.5truecm

{\bf Figure~1}:
Residuals of the best fit with a single blackbody 
(T$^\infty=1.0\times 10^6~^o$K, 
$N_H=5\times 10^{19}$ cm$^{-2},$  $\chi^2=2.20$ for 76 dof).
The presence of a hard tail above 1 keV is clearly visible.

\vskip 1.0truecm

{\bf Figure~2}:
Results of the fits with two component models: {\bf (a)} two blackbody
components;  {\bf (b)} sum of a  blackbody
and a power law.

\vskip 1.0truecm

{\bf Figure~3}:
Confidence contours (68\% and 90\% confidence level) for two interesting
parameters.
{\bf (a)} and {\bf (b)} refer to the fit with two blackbodies;
{\bf (c)} and {\bf (d)}   to the blackbody plus power law case.

\vskip 1.0truecm

{\bf Figure~4}:
Comparison between the PSPC light curve of PSR~0656+14 
in the energy range [0.10--2.40] keV (solid line) and that
calculated with the reference parameters of Table~\ref{tab:integ_par}
(dotted line). 
A typical error bar is shown.

\vskip 1.0truecm

{\bf Figure~5}:
Background subtracted light curves of PSR~0656+14 in three different
energy ranges. From top to bottom: 0.10--0.24 keV, 0.24--0.55 keV and
0.55--2.40 keV. Each curve has been normalized dividing by the average
number of counts per phase bin (respectively: 582, 579, 94 counts in
the low, medium and high energy range). The light curves have been
smoothed with a running average algorithm with five bins width. 

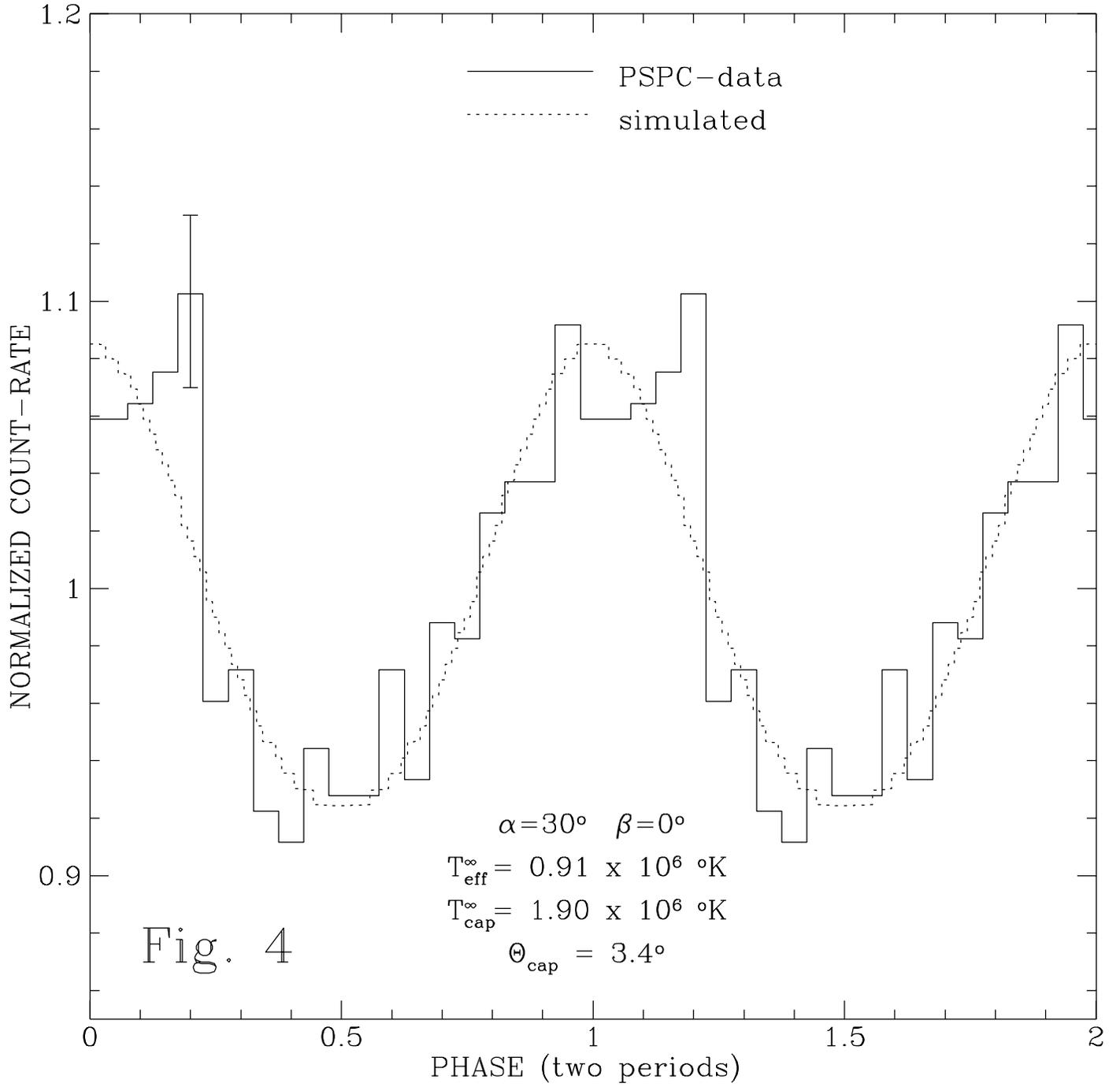
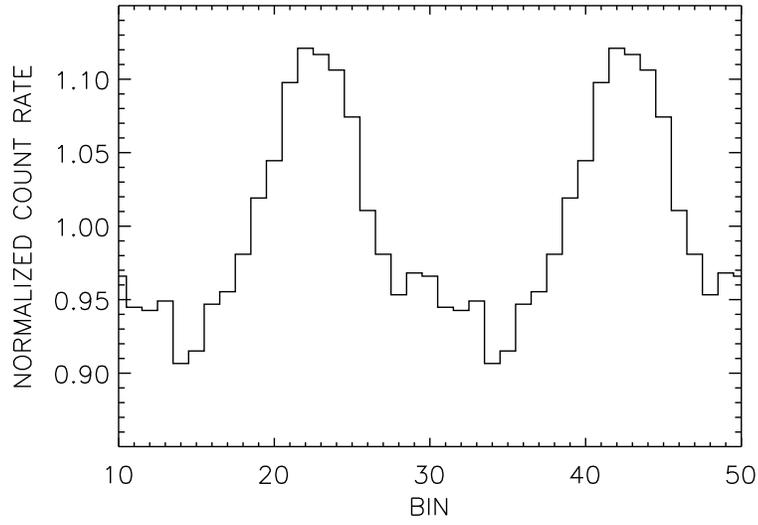
\end{document}